# Perspectives and Challenges in the Analysis of Prison Systems Data: a Systematic Mapping[1]

PERSPECTIVAS E DESAFIOS NA ANÁLISE DE DADOS DE SISTEMAS PRISIONAIS: UM MAPEAMENTO SISTEMÁTICO

*Glauco de Figueiredo Carneiro*[1]
*Rafael Antônio Lima Cardoso*[2]
*Antônio Pedro Dores*[3]
*José Euclimar Xavier de Menezes*[4]

1. DOUTOR EM CIÊNCIAS DA COMPUTAÇÃO, UNIFACS.
   https://orcid.org/0000-0001-6241-1612

2. MESTRANDO EM SISTEMAS E COMPUTAÇÃO, UNIFACS.
   https://lattes.cnpq.br/2712724638701758

3. DOUTOR EM SOCIOLOGIA, UNIVERSIDADE DE LISBOA (ISCTE-IUL).
   https://orcid.org/0000-0002-5482-6196

4. DOUTOR EM FILOSOFIA, UNIFACS.
   https://orcid.org/0000-0001-7839-7931

## ABSTRACT

**Context:** Open public data enable different stakeholders to perform analysis and uncover information from different perspectives. The identification and analysis of data from prison systems is not a trivial task. It raises the need for the research community to know how these data have been produced and used. **Goal:** Analyze prison systems data for the purpose of characterizing its use with respect to data sources, purpose and availability. **Method:** We performed a systematic mapping on existing evidence on prison systems original data from peer-reviewed studies published between 2000 and 2019. **Results:** Out of the 531 records, 196 articles were selected from the literature. **Conclusion:** The vast majority of the analyzed papers (75%) used restricted data. Only 18 studies (9%) provided data, which hampers replication initiatives. This indicates the need to analyze prison system in an integrated fashion, in which multidisciplinary and transparency are relevant issues to consider in such studies.

---

# Vulnerabilidades sociais convocam políticas públicas

## RESUMO

**Contexto:** Dados públicos abertos permitem que diferentes partes interessadas realizem análises e revelem informações de diferentes perspectivas. A identificação e análise de dados dos sistemas penitenciários não é uma tarefa trivial. Isso levanta a necessidade de a comunidade de pesquisa saber como esses dados foram produzidos e usados. **Objetivo:** Analisar os dados dos sistemas penitenciários com o objetivo de caracterizar seu uso quanto às fontes, finalidade e disponibilidade dos dados. **Método:** Foi realizado um mapeamento sistemático das evidências existentes sobre os dados originais dos sistemas penitenciários de estudos revisados por pares publicados entre 2000 e 2019. **Resultados:** Dos 531 registros, 196 artigos foram selecionados da literatura. **Conclusão:** A grande maioria dos artigos analisados (75%) utilizou dados restritos. Apenas 18 estudos (9%) forneceram dados, o que dificulta as iniciativas de replicação. Isso indica a necessidade de analisar o sistema prisional de forma integrada, em que a multidisciplinaridade e a transparência são questões relevantes a serem consideradas nesses estudos.

**Palavras-chave**: Sistema prisional; Dados abertos; Transparência; Transparência; Compartilhamento de informações de dados.

## 1 INTRODUCTION

Prisons are by definition places of detention for persons convicted and sentenced, whereas jails are used for pretrial and presentenced detainees (WENER, 2018). In this study, we use the word prison to refer to the place of detention as a reference to understand how researchers and stakeholders deal with data related to this subject. Total worldwide prison population has grown on considerable average that is similar to the estimated increase in the world's general population over the last ten years (WALMSLEY, 2014). Considering the representative percentage of people incarcerated, data related to the Prison Systems deserve attention by the research community for many reasons. The identification of trends in prison systems can be one relevant reason for public policy. For example, to which extent people with specific drug addiction have an increased incidence of incarceration. In this case, the possible association between incarceration and drug dependency should be analyzed, for example, among a cohort of injection drug users (KOEHN, BACH, *et al.*, 2015). On the other hand, researchers have argued the possibility of a specific enzyme degradation, as a key biological factor in the predisposition to impulsive aggression (STETLER, DAVIS, *et al.*, 2014). In this case, a representative sample of inmates is required to confirm this hypothesis.

# Vulnerabilidades sociais convocam políticas públicas

The data from Prison Systems enable different stakeholders, to perform analysis and uncover information from different perspectives, including in this case, the possibility of a cross-government information-sharing. This raises the need for the research community to know how these data have been used to accomplish aforementioned needs. Although studies analyzing prison systems data have widely adopted several types of perspectives and areas to focus the analysis, it lacks a proper understanding of how these individual studies contribute to the entire field of prison systems data. To the best of our knowledge, there is no secondary study that investigates how data related to Prison Systems´ have been used by the research community. For this reason, we conducted a Systematic Mapping Study (SMS) to gather evidence provided by papers published in peer-reviewed conferences and journals from January 2000 to May 2019. We initially found 509 papers as a result of the applied search strings in specific electronic databases and the execution of snowballing procedure (WOHLIN, 2014), from which we considered 196 studies as relevant. Findings suggest that there is a gap in effective solutions to deal with prison systems' data, especially data related to the prison system management. This indicates the need to motivate researchers to conduct studies in this subject. On the other hand, stakeholders need to be more sensitive to provide access to this type of data, considering privacy and ethics criteria.

This systematic mapping study is part of a larger joint project, which aims to propose a road map on how to identify, collect and analyze prison systems' data and an technological infrastructure to support the accomplishment of these goals. As a first step of this project, we endeavour to characterize how researchers from different domains have analyzed these data.

The remainder of this paper is organized as follows. Section 2 presents the design we adopted to conduct this systematic mapping study. Section 3 presents the key findings to the stated research questions and Section 4 discusses perspectives and challenges to conduct research analyzing prison systems data. Section 5 presents the threats to validity of our findings, and Section 6 concludes and presents ongoing work.

**2 RESEARCH DESIGN**

We performed a systematic mapping review of the published peer-reviewed literature to gather existing evidence on the use and analysis of data in prison settings from 2000 to 2019. In this study, we

CENTRO UNIVERSITÁRIO SOCIAL DA BAHIA – UNISBA   REVISTA DIÁLOGOS POSSÍVEIS | SALVADOR | V.20 N.1| JAN-JUN 2021

consider prison settings as prisons, jails and other custodial settings functioning as prison (excluding migrant centres and police detention rooms) (MADEDDU, VROLING, *et al.*, 2019). Systematic mapping is a type of secondary study that has the goal to describe the extent of the research in a field and to identify gaps in the research base. It identifies gaps in the research, where further primary research is needed, and areas where no systematic reviews have been conducted and there is scope for future review work (CLAPTON, RUTTER e SHARIF, 2009).

Systematic mapping provides descriptive information about the state of the art of a topic and a summary of the research conducted in a specific period of time (CLAPTON, RUTTER e SHARIF, 2009). The overall process for the selection of relevant studies is presented in Table 1 and described in more detail in the following subsections.

Table 1 - Steps for the Selection Process.

| **Step** | **Description** |
|---|---|
| (1) | Apply the search strings to obtain a list of candidate papers in specific eletronic databases. |
| (2) | Remove replicated papers from the list. |
| (3) | Apply the exclusion criteria in the listed papers. |
| (4) | Apply the inclusion criteria after reading abstracts, introduction and conclusion in papers not excluded in step 3. |
| (5) | Apply quality criteria in selected papers from step 4. |

**2.1 PLANNING**

We conducted this SMS based on a protocol comprised of objectives, research questions, selected electronic databases, search strings, and selection procedures comprised of exclusion, inclusion and quality criteria to select studies from which we aim to answer the stated research questions (WOHLIN e OTHERS, 2012).

The goal of this study is presented in Table 2 according to the Goal Question Metric (GQM) approach (BASILI e ROMBACH, 1988).

**Vulnerabilidades sociais convocam políticas públicas**

Table 2 - The Goal of this SMS according to the GQM Approach.

| Analyze | Prison Systems Data |
|---|---|
| *for the purpose of* | Characterization |
| *with respect to* | data sources, purpose, availability and area under analysis |
| *from the point of view of* | researchers and stakeholders |
| *in the context of* | both academia, official authorities, non-governmental bodies, civil society, organisations and policy makers |

The Research Question (RQ) is "How have researchers and stakeholders from academia, official authorities and non-governmental bodies characterized prison system's data regarding data sources, purpose, access permissions, availability and target domain in which data is analyzed based on papers published in the peer-reviewed literature"? The motivation behind RQ is justified by the acknowledgment that data sources, purpose of dealing with data, access permissions, availability and research focus are required to tackle issues or improvements related to the effective use of data in a given context (HEY, TANSLEY, *et al.*, 2009).

The specific research questions have the goal to gather evidence to support the answer of the stated RQ. This research question is in line with the goal of this review, and has been derived into four specific research questions, as follows. Specific Research Question 1 (SRQ1):

What are the target subjects that motivate data analysis in the context of prison systems by researchers and stakeholders from academia, official authorities and non-governmental bodies? This information enable us to identify main domain areas interested in prison systems data and to which extent they adopt a multidisciplinary approach in the data analysis. Specific Research Question 2 (SRQ2): What are the main data sources used in prison systems data studies? The data sources are a crucial component of any research project from which information can be extracted to unveil trends and patterns that otherwise would not be known by the research community. Specific Research Question 3 (SRQ3): What was the main goal of researchers and stakeholders from academia, official authorities and non-governmental bodies to deal with prison systems data? Considering that prison systems can be analyzed from different perspectives, there is the need to know how researchers deal with data: to perform analysis, to update data, to build data repositories or to replicate previous analysis. Specific Research Question 4 (SRQ4): What is the access permission of data related to prison systems available for researchers and stakeholders from academia, official authorities and non-governmental bodies? Prison systems have different types of data that need special attention in terms of security issues, including confidentiality, integrity, availability and authenticity.

The steps to build a search string to identify studies in the target repositories are shown in Tables 3 and 4. The Table 3 refers to major terms for the research objectives. We also considered the use of alternative terms and synonyms of these major terms. For example, the term prison systems can be associated with terms such as prison settings, jail and penal institutions. These alternative terms, as shown in Table 4, can be also included in the search string. We built the final search string by joining the major terms with the Boolean ``AND" and joining the alternative terms to the main terms with the Boolean ``OR". The focus of the formed search strings is to identify studies targeting the research questions of this systematic mapping.

The Research Question (RQ) is "*How have researchers and stakeholders from academia, official authorities and non-governmental bodies characterized prison system's data regarding data sources, purpose, access permissions, availability and target domain in which data is analyzed based on papers published in the peer-reviewed literature*"? The motivation behind RQ is justified by the acknowledgment that data sources, purpose of dealing with data, access permissions, availability and research focus are required to tackle issues or improvements related to the effective use of data in a given context (HEY, TANSLEY, *et al.*, 2009). The specific research questions have the goal to gather evidence to support the answer of the stated RQ. This research question is in line with the goal of this review, and has been derived into four specific research questions, as follows. Specific Research Question 1 (SRQ1): *What are the target subjects that motivate data analysis in the context of prison systems by researchers and stakeholders from academia, official authorities and non-governmental bodies?* This information enable us to identify main domain areas interested in prison systems data and to which extent they adopt a multidisciplinary approach in the data analysis. Specific Research Question 2 (SRQ2): *What are the main data sources used in prison systems data studies?*

# Vulnerabilidades sociais convocam políticas públicas

The data sources are a crucial component of any research project from which information can be extracted to unveil trends and patterns that otherwise would not be known by the research community. Specific Research Question 3 (SRQ3): *What was the main goal of researchers and stakeholders from academia, official authorities and non-governmental bodies to deal with prison systems data?* Considering that prison systems can be analyzed from different perspectives, there is the need to know how researchers deal with data: to perform analysis, to update data, to build data repositories or to replicate previous analysis. Specific Research Question 4 (SRQ4): *What is the access permission of data related to prison systems available for researchers and stakeholders from academia, official authorities and non-governmental bodies?* Prison systems have different types of data that need special attention in terms of security issues, including confidentiality, integrity, availability and authenticity.

The steps to build a search string to identify studies in the target repositories are shown in Tables 3 and 4. The Table 3 refers to major terms for the research objectives. We also considered the use of alternative terms and synonyms of these major terms. For example, the term *prison systems* can be associated with terms such as *prison settings*, *jail* and *penal institutions*.

These alternative terms, as shown in Table 4, can be also included in the search string. We built the final search string by joining the major terms with the Boolean ``AND'' and joining the alternative terms to the main terms with the Boolean ``OR''.

The focus of the formed search strings is to identify studies targeting the research questions of this systematic mapping.

Table 3 - Major terms for the research objectives.

| Criteria | Major Terms |
|---|---|
| (P)opulation | AND "prison system data" |
| (I)ntervention | AND ``approach`` AND ``analyze" AND "purpose" AND availability" AND ``type of data`` |
| (C)omparison | Not Applicable |
| (O)utcomes | AND ``analysis results`` |

# Vulnerabilidades sociais convocam políticas públicas

Table 4 - Alternative terms from major terms.

| Major Term | Alternative Terms |
|---|---|
| ``prison system'' | (``prison settings'' OR ``inmates'' OR ``custodial inmates'' OR ``incarcerated'' OR "prisoner" OR "jail" OR "penal institutions" OR "prison population" OR "Correctional Facility" OR "inmate locator" OR "Criminal Information" OR "custody information") |
| ``type of data'' | (``cohort'' OR ``data format'' OR) |
| ``availability'' | (``public data'' OR ``open access'' OR ``public record'' OR ``approach'' OR ``model'' OR ``methodology'' OR ``solution'') |
| ``purpose'' | (``usage`` OR ``utilization``) |

Table 5 presents the criteria for exclusion, inclusion and quality evaluation of papers in this review. The *OR* connective used in the exclusion criteria means that the exclusion criteria are independent, i.e., meeting only one criterion is enough to exclude the paper. On the other hand, the *AND* connective in the inclusion criteria means that all inclusion criteria must met to select the paper under analysis. Table 5 also presents the quality criteria used for this review represented as questions adjusted from their original version from Dyba and Dingsoyr.

(DYBA e DINGSOYR, 2008). We evaluated all the remaining papers that passed the exclusion and inclusion criteria using the quality criteria presented in the same table. All these criteria must met (i.e., the answer must be YES for each one) to permanently select the paper, otherwise the paper must be excluded. The exclusion, inclusion and quality criteria were used in the steps for the selection process as already presented in Table 1. According to Table 6, at the end of the selection process, all the retrieved papers were classified in one of the three options: *Excluded*, *Not Selected* and *Selected*.

Table 5 - Exclusion, Inclusion and Quality Criteria

| Type | Id | Description | Connective or Answer |
|---|---|---|---|
| Exclusion | E1 | Published earlier than 2000 | OR |
| Exclusion | E2 | The paper was not published in a peer-reviewed journal or conference | OR |
| Exclusion | E3 | The paper does not present primary nor secondary study | OR |
| Exclusion | E4 | The paper has less than 4 pages | OR |
| Inclusion | I1 | The paper must report an analysis of data related to prison systems | AND |
| Quality | Q1 | Are the aims of the study clearly specified? | YES/NO |
| Quality | Q2 | Is the context of the study clearly stated? | YES/NO |
| Quality | Q3 | Is the data analysis approach in line with the aims of the study? | YES/NO |

# Vulnerabilidades sociais convocam políticas públicas

## 2.2 EXECUTION

The quantitative evolution of the selection process execution is summarized in Figure 1.

The figure uses the PRISMA flow diagram (MOHER, LIBERATI, *et al.*, 2009) and shows the performed steps and the respective number of papers for each phase of the systematic mapping.

Table 6 - Classification Options for Each Retrieved Paper.

| Classification | Description |
|---|---|
| Excluded | Papers met the exclusion criteria. |
| Not Selected | Papers not excluded due to the exclusion criteria, but did not met the inclusion or quality criteria. |
| Selected | Papers did not meet the exclusion criteria and met both the inclusion and quality criteria. |

According to Table 1, as a result of the execution of Step 1 (execution of the search string), we retrieved from the three selected repositories a total of 509 papers (Identification Phase of Figure 1). The snowballing added 22 papers to the previous set, resulting in 531 papers. Considering that one paper was duplicated, we evaluated 530 regarding the alignment of their titles and abstracts to the stated specific research questions (Screening Phase of Figure 1).

The result of this evaluation was the exclusion of 316 papers and the inclusion of 214 papers, following the exclusion and inclusion criteria respectively already presented in Table 5. In the Eligibility Phase of Figure 1, we evaluated 214 papers to decide that 18 papers should have been not selected due to not meeting the three quality criteria presented in Table 5. The final set of studies to answer the specific research question is comprised of 196 papers (Included Phase of Figure 1).

**Vulnerabilidades sociais convocam políticas públicas**

Figure 1: Phases of the Selection Process in Numbers.

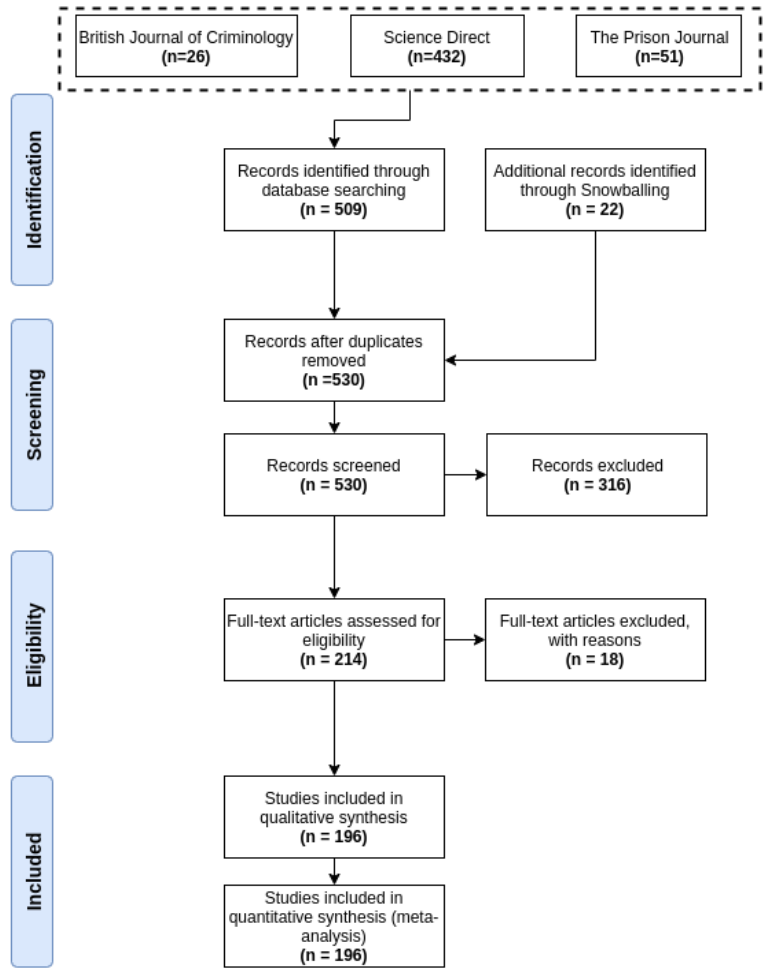

Table 7 presents the effectiveness of the search strings considering the 509 retrieved papers. The repository database that most contributed with selected studies was the *Science Direct* with 142 papers, corresponding to a search effectiveness of 32.8%.

The 174 selected studies represented 34.1% of all 509 papers retrieved by the search string. It is worth remembering that in this set of selected studies we added 22 studies obtained from snowballing (WOHLIN, 2014).

Table 7 - Effectiveness of the Search Strings.

| Database | Papers Retrieved by the Search String | Selected Papers | Search Effectiveness |
|---|---|---|---|
| Science Direct | 432 | 142 | **32.8 %** |
| The Prison Journal | 51 | 22 | **43.1 %** |
| The British Journal of Criminology | 26 | 10 | **38.4 %** |
| TOTAL | 509 | 174 | **34.1 %** |

# Vulnerabilidades sociais convocam políticas públicas

Figure 2: Evidence from the Literature to Answer Specific Research Questions.

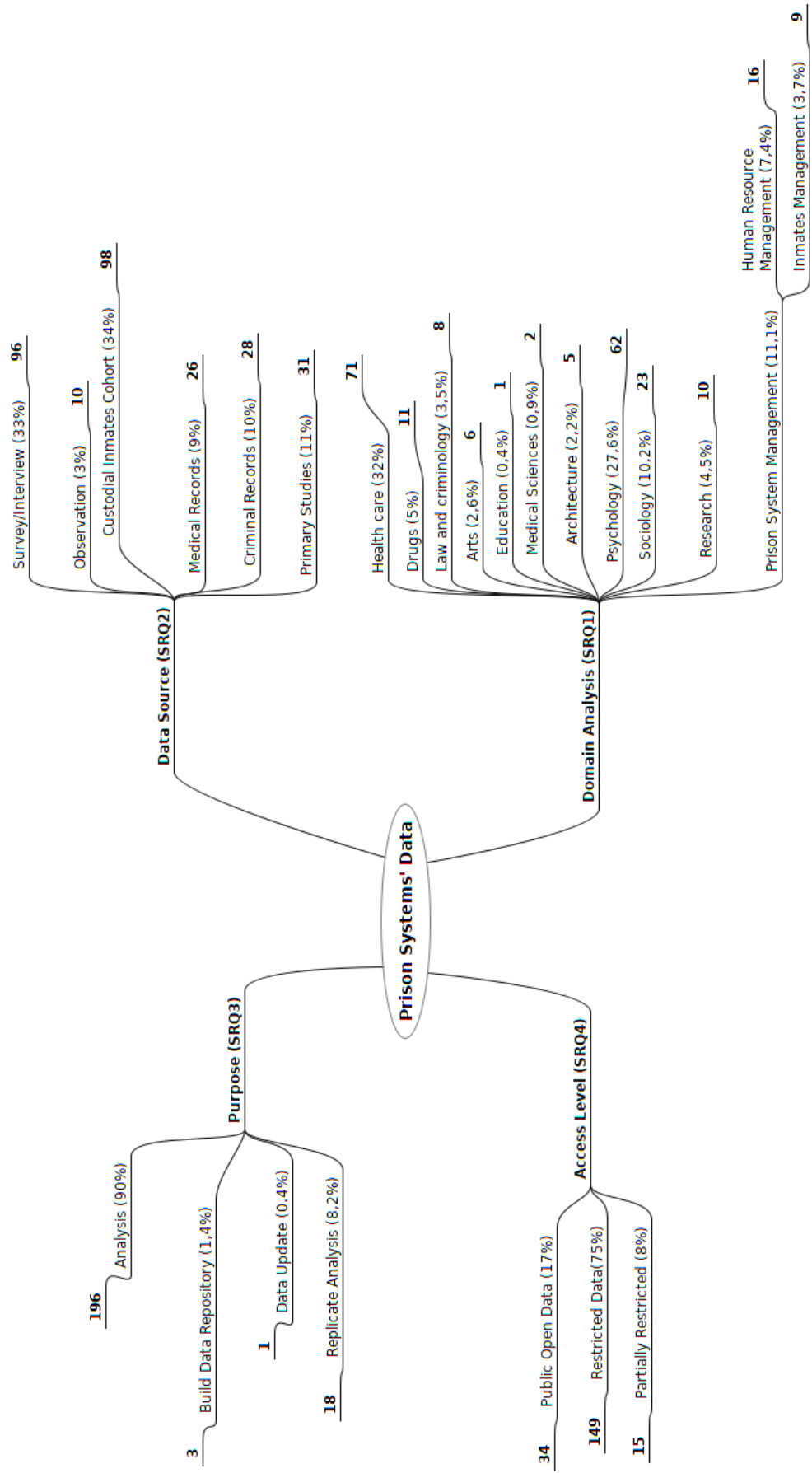

**Vulnerabilidades sociais convocam políticas públicas**

**3. RESULTS**

According to Ross and Tewksbury (ROSS e TEWKSBURY, 2018), the conduction of scholarly research on, in, and about correctional facilities is not a trivial task. The same authors list three main reasons for the inherent difficulty of such analysis (ROSS e TEWKSBURY, 2018).

First, considerable variability exists across jails, prisons, and correctional settings institutions and among the people that interact within these environments. Second, some types of facilities, correctional workers, and inmates are more likely to participate in the studies than others. This can significantly affect the characteristics of the analyzed sample.

Third, the access to these settings, the individuals who are both housed and work there and associated data is governed by stringent protocols that consider ethical, security and privacy issues, among others. Regarding the experience of researchers dealing with prison systems, there are very few articles published in major specialized journals and conferences that deal with the firsthand experiences of either prison staff or inmates (PATENAUDE, 2004).

Especially the requirements stated by procedural hoops, such as Institutional Review Boards (IRBs) at the researcher's home university or human subjects reviews can represent a relevant impediment to conduct research in correctional settings (PATENAUDE, 2004). After the grant of external approvals, researchers do still find a number of internal barriers to access the prison as a research site such as the need to obtain senior management approval of correctional settings (PATENAUDE, 2004).

Figure 2 presents evidence collected from the literature to answer the Specific Research Questions 1 to 4. Each branch has the associated Specific Research Question and the corresponding answers represented in sentences with the amount of studies from which they were collected. The total amount of references of each of the four branches of Figure 2 does not correspond to the total of 196 analyzed articles. The reason for this difference is that some studies fall in more than one the available groups.

**Answer to the Specific Research Question 1 (SRQ1)** — Figure 2 plots in the bottom right corner data related to the *Domain Analysis (SRQ1)* branch. We identified 11 key focus areas as follows: Health Care (71 studies — 32%), Drugs (11 studies — 5%), Law and Criminology (8 studies — 3,5%), Arts (6 studies —

CENTRO UNIVERSITÁRIO SOCIAL DA BAHIA – UNISBA    REVISTA DIÁLOGOS POSSÍVEIS | SALVADOR | V.20 N.1| JAN-JUN 2021

**Vulnerabilidades sociais convocam políticas públicas**

2,6%), Education (1 study — 0,4%), Medical Sciences (2 studies — 0,9%), Architecture (5 studies — 2.2%), Psychology (62 studies — 27,6%), Sociology (23 studies — 10,2%), Research (10 studies — 4,5%), and Prison Systems Management composed by Human Resource Management (16 studies — 7,4%), Infrastructure and Services Management and Inmates Management (2 studies — 3.4%).

The Health Care is by far the domain that attracted more studies. In the sequence, Psychology, Sociology and Human Resource Management stand out from the others in the amount of selected studies. One possible reason for this is that Health Care studies have availability of funding and scholarships when compared with others. Other possible reason is that prison settings have special conditions of a specific set of viruses and diseases that attract attention from Health Care researchers.

For example, human immunodeficiency virus (HIV) was discussed in S18, S22, S52, S78, S86, S89, S106, S119, S134, S137, S139, while hepatitis C virus (HCV) was discussed in S09, S14, S15, S94, S96. Regarding the multidisciplinary approach to analyze data, we identified studies that focused on more than one key area: S21, S46, S77, S131 (Health Care and Psychology), S37, S38 (Health Care and Drugs), S44 (Health Care and Law and Criminology), S165 (Architecture and Prison System Management).

A set of 10 studies discussed the conduction of research in prison settings. According to S191 (BROSENS, DE DONDER, *et al.*, 2015), despite the need for research in such environments, relatively few evaluations were conducted in prisons. In fact, the most cited among the selected paper presented in Table 11 has 238 citations, while this number could be considered as low when compared with most influential papers from other areas. Simpson et al. (SIMPSON, GUTHRIE e BUTLER, 2017) focused on the identification of research priorities and organizational issues in conducting research with prisoners, and ranking research priorities.

They invited prison health service directors in each Australian state and territory to participate in a national (deliberative) roundtable where the consensus building nominal group technique was utilized. Wener (S26) (WENER, 2018) argued that, unfortunately, prisons serve as an unintended laboratory to observe the influence of environmental conditions presented in levels that are unseen in other settings. Hence, in line with the same author, correctional environments are CENTRO UNIVERSITÁRIO SOCIAL DA BAHIA – UNISBA                    REVISTA DIÁLOGOS POSSÍVEIS | SALVADOR | V.20 N.1| JAN-JUN 2021

prone to profoundly affect the lives of those who reside there, as well as those who work in them. In this case, the author mentioned that not only the inmates are affected by the exceptional conditions, but also the correctional staff. This corroborates the relevance of the Prison System Management branch presented in Figure 2, corresponding to 11,2% of the studies and organized in the groups Human Resource Management and Inmates Management.

Higgins et al. (HIGGINS, TEWKSBURY e DENNEY, 2013), reported that the research on correctional staff stress has led to several themes, with regards to what aspects of their occupation they see as stress inducing, and the link between safety concerns that accompany their occupation and their perceived work stress.

**Answering the Specific Research Question 2 (SRQ2)** — Figure 2 plots data related to the Specific Research Question 2 (SRQ2) in the upper right branch. We classified the studies in the following types of data sources: Survey/Interview (33%), Observation (3%), Custodial Inmates Cohort (34%), Medical Records (9%), Criminal Records (10%) and Primary Studies (11%).

Likewise the rationale used to represent studies related to the SRQ1, a study can be grouped in more than one type, given that it can use different data sources. The data source classified as custodial inmate cohort refers to the cases in which researchers have authorized access to specific groups of inmates. Another important group of studies identified in the answer of SRQ2 is the data source classified as primary studies. We found out 31 secondary studies focusing on issues related to prison systems. This is an important finding, given that there are still researchers interested in analyzing peer-reviewed articles focusing on prison systems issues. However, this number is slow when compared to other well-known areas such as Software Engineering. For example, a tertiary study reported the analysis of 210 systematic mappings published in the literature focusing on software engineering issues (KHAN, SHERIN, *et al.*, 2019).

# Vulnerabilidades sociais convocam políticas públicas

Table 8 - Selected Studies grouped by Data Source (SRQ2).

| Data Source | Selected Studies |
|---|---|
| Survey/Interview (96) | S1, S3, S4, S9, S15, S16, S17, S18, S19, S20, S21, S29, S30, S31, S32, S33,S35, S39, S40, S41, S45, S49, S50, S51, S53, S55, S59, S60, S63, S66, S68, S69, S76, S77, S83, S84, S86, S90, S91, S93, S104, S105, S114, S115, S116,S118, S122, S123, S127, S129, S130, S134, S135, S136, S137, S138, S141, S142, S144, S145, S147, S148, S149, S150, S152, S154, S159, S161, S162, S165, S166, S169, S171, S172, S174, S175, S176, S177, S178, S179, S180, S181, S182, S183, S184, S185, S186, S187, S188, S189, S190, S191, S192, S193, S194, S196 |
| Observation (10) | S118, S138, S153, S155, S156, S157, S158,163, S164 |
| Custodial Inmates Cohort (98) | S1, S2, S4, S6, S8, S10, S13, S15, S16, S17, S23,S24,S26, S27, S32, S35, S36, S41, S42, S43, S44, S48, S49, S50, S51, S52, S55, S57, S58, S59, S60, S61,S62, S63, S64, S67, S74, S75, S78, S79, S80, S81, S83, S84, S86, S88, S89, S90, S92, S94, S95, S96, S98, S100, S101,S102, S103, S106, S108, S110, S111, S113, S114, S115, S116, S117, S118, S119, S120, S122, S123, S124, S125, S126, S127, S128, S129, S133, S134, S135, S136, S137, S139, S140, S141, S144, S145, S146, S147, S149, S152, S159, S161, S162, S169, S172, S187, S192 |
| Medical Records (26) | S1, S4, S8, S10 S13, S23, S24, S25, S36, S43, S54, S55, S57, S58, S64, S79, S82, S83, S97, S101, S109, S119, S120, S128, S139, 173 |
| Criminal Records (28) | S42, S44, S47, S48, S61, S62, S64, S65, S71, S72, S73, S82, S84, S87,S112, S120, S121, S125, S130, S141, S143, S146, S148, S150, S151, S160,S168, S173 |
| Primary Studies (31) | S5, S6, S7, S11, S12, S14, S19, S22, S26, S28, S34, S37, S38,S46, S52, S56, S70, S74, S77, S78, S85, S99, S107, S108, S117,S124, S126, S130, S131, S132, S143, S148, 151, S167, S168, S170,S177, S188, S189, S191 |

**Answering the Specific Research Question 3 (SRQ3)** — Figure 2 plots data related to the Specific Research Question 3 (SRQ3) in the upper left branch. Table 9 presents the selected studies that provide evidence to answer the Specific Research Question 3 (SRQ3).

The studies are classified in four groups according to the purpose of dealing with data: Analysis (90%), Build Data Repository (1.4%), Data Update (0.4%) and Replicate Analysis (8.2%). These findings indicate few initiatives to build data repository (3 studies) that is also consistent with the low replications reported in the selected studies (only 18 out of 196 studies).

**Answering the Specific Research Question 4 (SRQ4)** — Figure 2 plots in the bottom left corner the amount of selected studies distributed in the following data access levels groups: Public Open Data (17%), Restricted Data (75%) and Partially Restricted Data (8%). The complete list of studies that falls in each of these groups is presented in Table 10.

**Vulnerabilidades sociais convocam políticas públicas**

The evidence from these studies confirms that in most cases, the access to data is restricted and partially restricted, therefore, not public. The reason for this prevalence is that privacy and legal requirements must be guaranteed when dealing with personal data, including electronic medical records. Following this way, data stakeholders apply necessary safeguards to avoid information missuses. On the other hand, the need to apply for transparency, calls for public data available on this topic.

According to the authors, open data is viable when are related to aggregated and summarized data (BRAUNSCHWEIG, EBERIUS, *et al.*, 2012). This type of data is of common usage in secondary studies. On the other, we should bear in mind several initiatives to encourage the use of open data. In fact, open data offers many benefits to both researchers and practitioners. It increases the visibility of research results and encourages the reuse of data for new research questions and for verification purposes[1].

Table 9 - Selected Studies grouped by Purpose of Dealing with Data (SRQ3).

| Purpose of Dealing with Data (SRQ3) | Selected Studies |
|---|---|
| Analysis (196) | S1,S2, S3, S4, S5, S6, S7, S8, S9, S10, S11, S12, S13, S14, S15, S16, S17, S18,S19, S20, S21, S22, S23, S24, S25, S26, S27, S28, S29, S30,S31, S32, S33, S34, S35, S36, S37, S38, S39, S40, S41, S43, S44, S45, S46, S47, S48, S49, S50, S51, S52, S53, S54, S55, S56, S57, S58, S59, S60, S61, S62, S63, S64, S65, S66, S67, S68, S69, S70, S71, S72, S73, S74, S75, S76, S77, S78, S79, S80, S81, S82, S83, S84, S85, S86, S87, S88, S89, S90, S91, S92, S93, S94, S95, S96, S97, S98, S99, S100, S101, S102, S103, S104, S105, S106, S107,S108, S109, S110, S111, S112, S113, S114, S115, S116, S117, S118, S119,S120, S121, S122, S123, S124, S125, S126, S127, S128, S129, S130, S131, S132, S133, S134, S135, S136, S137, S138, S139, S140, S141,S142, S143, S144, S145, S146, S147, S148, S149, S150, S151,S152, S153, S154, S155, S156, S157, S158, S159, S160, S161, S162, S163, S164,S165,S166, S167, S168, S169, S170, S171, S172, S173, S174, S175, S176, S177, S178, S179,S180, S181, S182, S183, S184, S185, S186, S187, S188,S189, S190, S191,S192, S193, S194, S195,S196 |
| Build Data Repository (3) | S1, S2, S6 |
| Data Update (1) | S1 |
| Replicate Analysis (18) | S6, S10, S13, S20, S22, S23, S31, S38, S47, S67, S68, S79, S83, S87, S122, S150, S157, S195 |

---

[1] https://www.openaccess.nl/en/events/national-plan-open-science-presented

**Vulnerabilidades sociais convocam políticas públicas**

Table 10 - Selected Studies grouped by Access Level (SRQ4).

| Access Level(SRQ4) | Selected Studies |
|---|---|
| Public Open Data (34) | S15, S16, S17, S19, S24, S26, S34, S36, S37, S46, S52, S56, S70, S71, S72, S74, S79, S82, S85, S99, S107, S108, S124, S125, S126, S131, S132, S153, S155, S156, S163, S167, S168, S189, S195 |
| Restricted Data (149) | S1, S2, S3, S4, S5, S8, S9, S10, S13, S18, S20, S21, S23, S24, S25, S27, S29, S30, S31, S32, S33, S35, S39, S40, S41, S43, S44, S47, S48, S51, S53, S55, S57, S58, S59, S60, S61, S62, S63, S64, S65, S66, S67, S68, S69, S70, S71, S73, S75, S76, S77, S78, S80, S81, S83, S84, S85, S86, S87, S88, S89, S90, S91, S92, S93, S94, S95, S96, S97, S98, S99, S100, S101, S102, S103, S104, S105, S106, S109, S110, S111, S112, S113, S114, S115, S116, S117, S118, S119, S120, S121, S122, S123, S127, S128, S129, S133, S134, S135, S136, S137, S138, S139, S140, S141, S142, S143, S144, S145, S146, S147, S148, S149, S150, S151, S152, S154, S157, S158, S159 S160, S161, S162, S164, S165, S166, S169, S170, S171, S172, S173, S174, S175, S176, S177, S178, S179, S180, S181, S182, S183, S184, S185, S186, S187, S188, S191 S192, S193, S194, S196 |
| Partially Restricted (15) | S5, S6, S7, S11, S12, S14, S22, S28, S38, S42, S45, S49, S71, S130, S190 |

Table 11 - Most Influential Studies.

| ID | Year | Title | Domain Analysis | No. Citations |
|---|---|---|---|---|
| S119 | 2007 | HIV in prison in low-income and middle-income countries | **Health Care** | **238** |
| S113 | 2009 | Randomized Controlled Pilot Study of Cognitive-Behavioral Therapy in a Sample of Incarcerated Women With Substance Use Disorder and PTSD | **Psychology** | **193** |
| S142 | 2001 | Job satisfaction among detention officers: Assessing the relative contribution of organizational climate variables | **Prison System Management** | **167** |
| S125 | 2006 | he correctional melting pot: Race, ethnicity, citizenship, and prison violence | **Sociology** | **159** |
| S124 | 2006 | Meticillin-resistant Staphylococcus aureus among US prisoners and military personnel: review and recommendations for future studies | **Health Care** | **156** |

## 4. PERSPECTIVES AND CHALLENGES IN THE ANALYSIS OF PRISON SYSTEMS DATA

The answers to the specific research questions discussed in Section 3 provided an update and comprehensive overview which jointly account for the answer of the main research question. In this section, we discuss perspectives and challenges related to the main research question aiming at characterizing prison system's data regarding data sources, purpose of dealing with data, access permissions, availability and target domain.

## **Vulnerabilidades sociais convocam políticas públicas**

*Challenges related to the Domain of Analysis.* The interest for the Health Care domain is illustrated in Figure 2 by the preference of 32% of the selected studies. In this domain, themes vary among prevention of infect contagious diseases, harm reduction programs for drug and alcohol users, cancer detection and prevention, vaccination programs, supporting sexual abuse victims, neurological and psychological assessments, among others. One of the possible reasons for this attention from the research community is the structural persistence of symptoms of diseases coming out of prison facilities that has direct impact on public health. The main challenge is that these investigations require the participation of inmates to provide data to the study.

*Perspectives related to the Domain of Analysis.* Such an example of search for answers to avoid health care problems, such as avoidable loss of lives both inside and outside prisons, shows the advantage of data available for production to researchers and others. Healthcare data can be used together with criminal data and/or prison data. For example, observing and characterizing prison activities provide conditions to learn about stressed and on edge situations. Unfortunately, diseases can not be contained by walls. By its own sake, all society should be concerned with that part of life in prison. Art and education represent the philanthropic and professional concern on helping inmates to get in touch with social reality, through mind and manual work, to easy prison management and social reintegration. Another interesting perspective, would be the motivation to conduct multidisciplinary studies combining different sources of information ranging from social work services, from police, from healthcare histories, from prison observation of inmates, to get a comprehensive understanding of a specific situation and/or condition. They can also develop comparative or and longitudinal studies about prison facilities or about inmates.

The recorded domains of analysis of the selected papers indicate few law and criminology and management studies. Therapeutic perspectives are dominant: health care and psychology. Also, sociology is represented in our sample, probably also using a therapeutic approach to prisoners, not a management approach or a criminal approach. Most findings from management studies are related to human resources approach regarding prison officers. We have noticed that there are very few initiatives to link the professional and the trusty populations, even the main management problem of prison systems is the relationship between these two populations. Approaches to prison studies

CENTRO UNIVERSITÁRIO SOCIAL DA BAHIA – UNISBA                                REVISTA DIÁLOGOS POSSÍVEIS | SALVADOR | V.20 N.1| JAN-JUN 2021

that consider prisoners do not consider prison officers. For instance, architectural studies on prisons buildings focus only on inmate's needs. The study of these needs are not supported by health care, psychology studies and security management. We presume that digging in the structure of prison data will show how to rationalize public access to more data than the information available today, giving new opportunities to stakeholders research their own study subjects easily, cheaper, more times, and in a more accurate way. The conduction of studies that consider both prisoners and prison officers will certainly bring new insights for an effective comprehension of prison systems data.

*Challenges related to Open Data Access and Data Granularity.* Considering that storage and update practices (answer to the Specific Research Question 3 — SRQ3) are required to open access initiatives, evidence from the 196 selected studies indicate that most researchers do not meet this requirement. Almost every selected study explores data available in third part repositories or obtain data from surveys/interviews (33%), custodial inmates cohort (34%), medical and criminal records (9% and 10%, respectively), as already discussed in the answer to the Specific Research Question 2 (SRQ2). In fact, researchers have argued that while opening data has important benefits, sharing data comes with inherent risks to privacy and security issues (GREEN, CUNNINGHAM, *et al.*, 2017). Granularity is a relevant theme in this discussion, given that it refers to the level of detail of the units of data available in a repository. The low level of granularity contains high level of detail while the high level of granularity contains low level of detail (INMON, 2005).

At the heart of this dilemma lie two traits of granularity to open data: benefit (utility) and risk (privacy). These two attributes are often in conflict. In one hand, less data granularity implies in protecting privacy. On the other hand, more data granularity provides conditions as an asset to promote transparency, enable innovation, and aid research to unveil trends, gaps and improvement opportunities that without proper granularity would not be viable (GREEN, CUNNINGHAM, *et al.*, 2017). In the case of prison systems data, there is a critical trade-off between open data (transparency) to provide data granularity due to the details and restricted data (privacy). As a result of non-adherence to open data initiatives, very few of the selected studies (18 out of 196) provide data for replication, as can be seen in Table 12.

**Vulnerabilidades sociais convocam políticas públicas**

Table 12 - Selected Studies that Provide Data for Replication.

| Domain Analisys(SRQ1) | Data Available for Replication |
|---|---|
| Health Care (7) | S10, S20, S22, S23, S38, S67, S83 |
| Arts (1) | S157 |
| Psychology (7) | S6, S13, S31, S68, S79, S87, S122 |
| Sociology (2) | S47, S150 |
| Prison System Management (1) | Inmates Management - S195 |

*Perspectives related to Open Data Access and Data Granularity*. We have found initiatives that provide prison systems open data as can be seen in the datasets available in Kaggle, a public data platform [2]. We identified 16 public datasets that contain data related to prison systems. None of them were cited in the 196 selected studies of this systematic mapping. These datasets have different types of data related to the theme and different data granularity. For example, the State of New York (USA) hosts open datasets covering topics that range from farmers' markets to solar photovoltaic projects [3]. Among these topics, there is a dataset related to prison systems that represents inmate admissions to the NYS Department of Corrections and Community Supervision for a new offense or for a parole violation by month of admission [4]. The dataset includes data about admission type, county, gender, age, and crime. This type of dataset do not only provide data, they enable researches to analyze them through appropriate programming techniques according to their research goals. This is a viable form to provide data for research purposes in the context of prison systems.

*Challenges related to Data Sources*. We identified two main data sources of preference by researchers: official data and new data produced by each research process, mostly surveys (33%, according to Figure 2). In some cases, for medical proposes, the research build a set of biological samples collected from inmates. Official data used can be supranational or national stats or prison facilities´ data both from inmates. Prison facilities can also dispose of administrative and financial data useful to accomplish research goals. Researches use information to design diseases or violence diagnosis, to plane representative samples of inmates, for instance. Unfortunately, this data is not organized and structured for use. Moreover, the access to this type of data source depends on authorization. The administrative processes to get the authorization can take long, when successful. Another challenge is the format of data that can be digital or in paper (hard copy). Digital data requires much less effort to be processed and used by different stakeholders than in non-digital format. Digital data source can include audio, video, files and bio-metric data such as fingerprint, face recognition, DNA, iris recognition, among others.

---

[2] https://www.kaggle.com/datasets?search=prison
[3] https://data.ny.gov/}.
[4] https://www.kaggle.com/new-york-state/nys-prison-admissions-beginning-2008

*Perspectives related to Data Sources.* We argue that to receive the best collaborative information from inmates is to show everybody that everyone can be sure that the data would be used for benign and social, not manipulative and secret, proposes. One of the ways to develop confident and everyone stimulate collaboration to the prison´s data production and use processes would be to design open and collaborative information systems. This would be a transparent initiative prone to produce, at the same time, available data and public confidence. For this reason, a new area of main concern on prison studies should be data production, storage, accessibility and management.

Table 13 - Selected Studies with Financial Support.

| Domain Analisys(SRQ1) | Financial Support |
|---|---|
| Health Care (35) | S4, S5, S8, S9, S10, S11, S14, S18, S19,S20, S21, S22, S33, S35, S36, S38, S52, S53, S55, S67, S75, S78, S84, S99, S104, S108, S114, S116, S119, S124, S128, S129, S135, S162 |
| Drugs (7) | S59, S63, S95, S106, S136, S137, S152 |
| Law And Criminology (4) | S56, S112, S148, S149 |
| Psychology (27) | S13, S16, S24, S27, S29, S30, S31, S32, S39, S54, S58, S64, S65, S68, S69, S79, S87, S90, S91, S98, S101, S113, S115, S118, S127, S144, S170 |
| Sociology (12) | S12, S45, S51, S61, S62, S71, S81, S102, S143, S146, S150, S196 |
| Research(3) | S188, S190, S191 |
| Prison System Management (7) | Human Resource Management (4) - S142, S151, S175, S178 |
| | Inmates Management (3) - S190, S191, S196 |

Table 14 - Secondary Studies.

| Domain Analisys | ID |
|---|---|
| Health Care (18) | S5, S7, S11, S14, S19, S22, S34, S37, S38, S46, S52, S85, S99, S108, S124, S126, S131, S132 |
| Drugs (2) | S37, S38 |
| Law And Criminology (1) | S56 |
| Architecture (1) | S26 |
| Psychology (9) | S6, S26, S28, S46, S70, S74, S107, S117, S131 |
| Sociology (1) | S12 |
| Research (2) | S26, S189 |
| Prison System Management (4) | S130, S131, S167, S195 |

## 5 THREATS TO VALIDITY

We followed the guidelines proposed by (CLAPTON, RUTTER e SHARIF, 2009) (WOHLIN e OTHERS, 2012) to conduct this systematic mapping. However, we identified some threats to the validity of this study to be discussed in this section. The first threat might be the incompleteness of the selected studies that depends upon the limitations of the search engines and the keywords used in the search string. We adjusted the search string based on suggestions of specialists in the subject. Moreover, we also used synonyms (alternative terms) to build the search strings and therefore to eliminate non-relevant primary studies. We also made use of snowballing, along with database search, to add reliability to the review through the inclusion of studies targeting issues related to the research questions of this systematic mapping. Furthermore, the outcomes of the data extraction steps presented in Table 1 were confirmed by invited researchers. We managed to eliminate the search bias as much as feasible by adopting the guidelines and defining selection criteria as discussed in Subsection 2.1 and presented in Table 5.

We limited the search in repository engines that are related to academic research. Other studies published as non-academic books and grey literature, such as technical reports, white papers, work in progress, were not included in this study. Although we recognize that additional relevant published studies may have been overlooked, we believe that despite that limitation, this systematic mapping provides a relevant contribution to the discussed subject.

**Vulnerabilidades sociais convocam políticas públicas**

## 6. CONCLUSION

Imprisoning people is the most drastic sanction a society can impose on its citizens (WENER, 2018). For this reason, and considering the relevance of the theme as whole, prison settings deserves attention by the research community. Considering this scenario, we searched for published papers in the peer-reviewed papers to present a panoramic view of how the literature has dealt with issues related to data in correctional settings.

The findings discussed in this systematic mapping is a first attempt to close the gap of the absence of effective data science strategies between the nowadays situation of prison data and a new scenario in which data would be available to whom it may concern, without the risk of violation of legitimate secrecy deserved both by individuals and organizations.

The evidence provided by the selected studies also showed few public discussions about technology and methodologies to deal with prison systems data. The type of usage from the majority of selected studies was related to analysis of collected data much more than data storage and update. Data sources are mostly produced ad-hoc by each research group, which means lots of efforts to produce data for private and episodic use only.

We conclude that there is a potential for more secondary studies in prison systems area. As future work, we aim at analyzing the 31 secondary studies and others published in the literature to identify the areas of prison systems most discussed in these studies. The areas that lack secondary studies, such as data related to prison systems analyzed in the study.

## REFERENCIES

R. E. Wener, Can correctional environments be humane? a case for evidenceand value-based design(S26), in: Environmental Psychology and Human Well-430Being, Elsevier, 2018, pp. 281–311.

R. Walmsley, World pre-trial/remandimprisonment list, World prison popu-lation: International centre for prisonstudies.435

J. D. Koehn, P. Bach, K. Hayashi, P. Nguyen, T. Kerr, M.-J. Milloy, L. Rieb, E. Wood, Impact of incarcer-ation on rates of methadone use in acommunity recruited cohort of injection440drug users(S63), Addictive Behaviors 46(2015) 1 – 4.

D. A. Stetler, C. Davis, K. Leavitt, I. Schriger, K. Benson, S. Bhakta, L. C. Wang, C. Oben, M. Watters, T. Hagh-

445negahdar, M. Bortolato, Association oflow-activity maoa allelic variants withviolent crime in incarcerated offenders(S79), Journal of Psychiatric Re-search 58 (2014) 69 – 75.450

C. Wohlin, Guidelines for snowballing insystematic literature studies and a replication in software engineering, in: Proceedings of the 18th international con-23 ference on evaluation and assessment455in software engineering, Citeseer, 2014,p. 38.

G. Madeddu, H. Vroling, A. Oordt-Speets, S. Babudieri, ́E. O'Moore, M. V.Noordegraaf, R. Monarca, P. L. Lopalco,460D. Hedrich, L. Tavoschi, Vaccinationsin prison settings: A systematic reviewto assess the situation in eu/eea coun-tries and in other high income countries(S11), Vaccine.465

J. Clapton, D. Rutter, N. Sharif, Sciesystematic mapping guidance, London:SCIE.

C. Wohlin, et al., Experimentation inSoftware Engineering, Springer-Verlag,4702012.

V. R. Basili, H. D. Rombach, The tameproject: towards improvement-orientedsoftware environments, IEEE Transactions on Software Engineering 14 (6)475(1988) 758–773.

A. J. Hey, S. Tansley, K. M. Tolle, et al.,The fourth paradigm: data-intensive scientific discovery, Vol. 1, Microsoft research Redmond, WA, 2009.480

T. Dyba, T. Dingsoyr, Empirical studiesof agile software development: A systematic review, Information and SoftwareTechnology 50 (9) (2008) 833 – 859.

D. Moher, A. Liberati, J. Tetzlaff, D. G.485Altman, P. Group, et al., Preferred reporting items for systematic reviews andmeta-analyses: the prisma statement,PLoS medicine 6 (7) (2009) e1000097.

J. I. Ross, R. Tewksbury, The challenges490of conducting research on supermax prisons: Results from a survey of scholarswho conduct supermax research(S189),The Prison Journal 98 (6) (2018) 722–737.495

A. L. Patenaude, No promises, but i'mwilling to listen and tell what i hear:Conducting qualitative research amongprison inmates and staff(S188), ThePrison Journal 84 (4suppl) (2004) 69S–50091S.

D. Brosens, L. De Donder, S. Dury,D. Vert ́e, Building a research partnershipin a prison context: From collaborationto co-construction(S193), Sociological505Research Online 20 (3) (2015) 1–15.

P. L. Simpson, J. Guthrie, T. Butler,Prison health service directors' viewson research priorities and organiza-tional issues in conducting research in 510 prison: outcomes of a national deliber-ative roundtable(S190), Internationaljournal of prisoner health 13 (2) (2017)113–123.

G. E. Higgins, R. Tewksbury, A. S. Den-515ney, Validating a measure of work stressfor correctional staff: A structural equation modeling approach(S186), Criminal Justice Policy Review 24 (3) (2013)338–352.520

M. U. Khan, S. Sherin, M. Z. Iqbal,R. Zahid, Landscaping systematic mapping studies in software engineering: Atertiary study, Journal of Systems andSoftware 149 (2019) 396–436.52524

K. Braunschweig, J. Eberius, M. Thiele,W. Lehner, The state of open data, Limits of current open data platforms.

B. Green, G. Cunningham, A. Ekblaw,P. Kominers, A. Linzer, S. P. Crawford,530Open data privacy, Berkman Klein Center Research Publication (2017-1) (2017)17–07.

W. H. Inmon, Building the data warehouse, John wiley & sons, 2005.

# Vulnerabilidades sociais convocam políticas públicas

**Appendix A**

Avaliable at:

https://doi.org/10.5281/zenodo.3955201

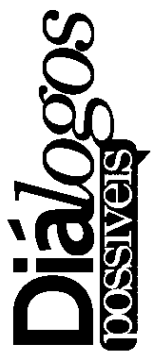

*REVISTA DIÁLOGOS POSSÍVEIS*
**Editor:** Professor Doutor José Euclimar Xavier Menezes

Centro Universitário Social da Bahia (UNISBA)

Avenida Oceânica 2717, CEP – 40170-010
Ondina, Salvador – Bahia.

**E-mail:** dialogos@unisba.edu.br
**Telefone:** 71- 4009-2840